\documentstyle[art12]{article}
\def\pt{p_{\bot}}

\def\mc{m_c}
\def\y{y^{\star}}
\def\mt{M_{\bot}}
\def\g{\gamma}
\def\sq{\sqrt{s}}
\def\cg{c(\gamma)g\to cg}
\begin{document}
\begin{center}
\bf {CHARM PHOTOPRODUCTION AT LARGE TRANSVERSE MOMENTUM\\
AND THE CHARM
CONTENT OF THE PHOTON}\\
V.A.~Saleev\\
Samara State University, Samara, 443011, Russia
\end{center}

\begin{abstract}
The charm quark photoproduction at large transverse momentum in the
resolved photon interaction via partonic subprocess $\cg$ is studied.
It is shown that at HERA energies the contribution of the charm quark
excitation in the photon in the inclusive charm production at $\pt>5$
GeV/c dominates over the other resolved photon contribution via the
gluon content of the photon and it is about 30\% of the contribution
of the $c\bar c$-pair production via the photon-gluon fusion mechanism.
\end{abstract}

The study of the charm quark photoproduction on protons at high energies is a
clean test of QCD as well as our knowledge of the gluon distribution
in the proton \cite{1}-\cite{6}. Recent experiment at HERA ep-collider have
shown that the resolved photon interaction, which probes the photon
structure, also can be very important in the charm quark
photoproduction process \cite{7}.

It is well known that the pure perturbative QCD calculations are not
adequate to describe the observed charm photoproduction data for $\pt$
and $x_F$ distributions. There are strong dependence on the value of
the c-quark mass, $k_{\bot}$-smearing and other nonperturbative
effects. In the region of large transverse momentum $\pt>>\mc$ the
results based on the perturbative QCD must be more reliable. Previous
calculations of the c-quark inclusive photoproduction at large $\pt$
have included LO and NLO direct charm production via photon-gluon
fusion \cite{1}-\cite{6}, including the semihard approach calculations
\cite{2,8,9,10}, the resolved photon contribution via subprocess
$g(\gamma)g\to c\bar c$ \cite{5,11} and the  diffractive-like photoproduction
in the vector-meson-dominance (VMD) model \cite{12}.
Recently we have shown the important role of the c-quark photon
structure function (PSF) in the $J/\psi$ \cite{13} and $B_c$-meson
\cite{14}
photoproduction at large transverse momentum via partonic subprocesses
$c(\gamma)g\to J/\psi c$ and $c(\g)\g\to B_c b$.
It is obviously that the
similar mechanism also plays a crucial role in the open charm
photoproduction at large $\pt$ via subprocess $\cg$. Note, that
early the charm quark hadroproduction via subprocess $cg\to cg$
\cite{15}
as well as $J/\psi$ hadroproduction via subprocess $cg\to J/\psi c$
\cite{15b} have
been studied, based on charm quark excitation model.

The conception of the intrinsic charm quark in the photon or the
proton is theoretically justified in the processes, where the charm
quark in the initial parton state receives sufficient transfer momentum
$\hat t$, which is necessary to excite $c\bar c$ pair. This condition
takes place for charm quark photoproduction at the large $\pt$, where
the relevant scale is $Q^2\sim \mt^2=\pt^2+\mc^2>>\mc^2$.

In the parton model the measurable cross section for the charm quark
photoproduction is defined by the parton level QCD cross section
$\hat\sigma(cg\to cg)$ and the respective parton distribution functions:

\begin{eqnarray}
\frac{d\sigma}{d\pt^2d\y}&&(\gamma p\to cX)= \nonumber\\
&&
\int_{x_{1,min}}^1dx_1 \frac{C_{\g}(x_1,Q^2)G_p(x_2,Q^2)}
{x_1-\frac{\mt}{\sqrt{s}}e^{\y}} \frac{d\hat\sigma}{d\hat t}(cg\to cg),
\end{eqnarray}
where
$$x_2=\frac{x_1\sqrt{s}\mt e^{-\y}-2\mc^2}{x_1 s-\sqrt{s}\mt e^{\y}},
\qquad x_{1,min}=\frac{\sqrt{s}\mt e^{\y}-2\mc^2}{s-\sqrt{s}\mt e^{-\y}},$$
$\hat t=2\mt^2-x_1\sqrt{s}\mt e^{-\y}$, $\hat u=\mt^2-x_2\sqrt{s}\mt
e^{\y}$, $\hat s=\mc^2+x_1x_2s$, $\y$ and $\pt$ are the c-quark rapidity
and transverse momentum, $s$ is
the square of a total energy in $\gamma p$ center of mass reference
frame. The explicit formula for the differential cross section
$\frac{d\hat\sigma}{d\hat t}(gc\to gc)$ is taken from \cite{15}. All
our calculations are based on the LO-GRV gluon distribution in the proton
$G_p(x,Q^2)$ \cite{16} evaluated at the scale $Q^2=\mt^2$. For c-quark
PSF we use phenomenological parameterization based on the VMD model
\cite{12} as well as QCD motivated LO-GRV parameterization \cite{17}.
In the Fig.1 the x-dependence of the c-quark PSF at the scale
$Q^2=4\mc^2$ is presented for both parameterizations. In
spite of the different x-dependence, both parameterization gives the
approximately equal values for the mean value of the photon momentum,
which is carried by charm quarks: $\sim 0.7\cdot 10^{-3}$.

The results of the calculation of the $\pt-$spectra for c-quark
photoproduction at $\sqrt{s}=200$ GeV are shown in Fig.2. The curve 1
is the contribution of the photon-gluon mechanism. The curves 2 and 3
are contributions of the c-quark PSF at the different values of the
dynamical cutoff \cite{15}: curve 2 corresponds to $|\hat
t_{min}|=\mc$ and curve 3 -- $|\hat t_{min}|=4\mc^2$. One has strong
suppression for the c-quark PSF contribution at small $\pt$ versus
$|\hat t_{min}|$. Howere, the results of the calculation are
independent of $|\hat t_{min}|$ at the large $\pt>5$ GeV/c, where the
condition $Q^2=\pt^2+\mc^2>>\mc^2$ is satisfied. The curve 4 in Fig.2
is the contribution of the gluon PSF via subprocess $g(\g)g\to c\bar
c$. We can see that the contribution of the gluon PSF is large smaller
than the contribution of the c-quark PSF at all $\pt>5$ GeV/c. This
conclusion is independent of the c-quark PSF parameterization.

Fig.3 shows the $\y$-spectra for c-quark photoproduction at
$\sqrt{s}=200$ GeV and $\pt>5$ GeV/c. The curves 1 and 4 are the same
as in Fig.2. The curves 2 and 3 are the contributions of the c-quark
PSF at the different parameterization of the PSF: 2 is GRV
parameterization \cite{17}, 3 is VMD parameterization \cite{12}.
The sufficient difference between curve 2 and 3 gives us opportunity
to check experimentally the nature of the charm content of the photon
in the c-quark photoproduction at the large $\pt$. The region of
$\y<0$ is also sensitive to the gluon PSF contribution. At $\y<0$ the
total contribution of the resolved photon interaction ( gluon plus
c-quark PSF) becomes larger than the contribution of the direct
photon-gluon fusion mechanism (curve 1).

In the Fig.4 the results of the calculation for the total cross
section of the charm quark photoproduction at $\pt>5$ GeV/c and all
$\y$ are presented. The notation as in Fig.3. The contribution of the
gluon PSF is strongly suppressed to compare with the c-quark PSF
contribution. The last one is about 30\% from the dominant photon-gluon
fusion contribution at $\sqrt{s}=200$ GeV and beyond. The both
parameterizations of the c-quark PSF gives the equal contributions,
which speedily growth versus energy beginning with $\sqrt{s}=50$ GeV.

In conclusion we want note that our analyze shows the
important role of the charm content in the photon for large $\pt$
charm quark photoproduction at HERA energy and beyond, independently
on the choice of the c-quark PSF parameterization.
Our calculation, based on PSF approach, sums up the large logarithmic
terms $log(\pt^2/m^2)$ and gives more reliable predictions for large
$\pt$ charm photoproduction than the pure NLO perturbative QCD
calculations of the $2\to 3$ subprocesses.
When our work was
near to completion, a related study appeared in ref.\cite{18}. Taking
into account that we use different sets of the parton densities in the
proton and the photon, our conclusions in general is the same as in
\cite{18}.

  This  research was supported by the Russian
  Foundation of Basic Research (Grant 93-02-3545
  and by State Committee on High Education of Russian Federation
  (Grant 94-6.7-2015). Author great thank N.Zotov and A.Martynenko for
useful discussions the problems of the photon structure function and
heavy quark photoproduction at high energy.

{\bf Figure captions}

\begin{enumerate}
\item The charm quark distribution in the photon at the scale
$Q^2=4\mc^2$. The curve 1 corresponds to the parameterization
\cite{12}, the curve 2 - \cite{17}.
\item The $\pt$ distribution for c-quark photoproduction at $\sq=200$ GeV
and all $\y$. The curve 1 is the direct photon-gluon fusion
contribution. The  curve 4 is the resolved photon
contribution via the $g(\g)g\to c\bar c$ subprocess. The short-dashed curves
are the
contribution of the charm quark excitation in the photon, the
curve 2 corresponds to $|\hat t|_{min}=\mc^2$ and the  curve 3
corresponds to $|\hat t|_{min}=4\mc^2$.
\item The $\y$-distribution for the c-quark photoproduction at $\sq=200$ GeV
and $\pt>5$ GeV/c. The curve 1 is the direct photon-gluon fusion
contribution. The curve 4 is the resolved photon contribution via the
$g(\gamma)g\to c\bar c$ subprocess. The short-dashed curves are the
contributions of the c-quark PSF, the curve 2 corresponds to VMD
parameterization \cite{12} and the curve 3 corresponds to GRV
parameterization \cite{17} of the c-quark PSF.
\item The total cross section for the c-quark photoproduction
 at all $\y $ and $\pt>5$ GeV/c versus $\sq$. Notation as in Fig.3.
\end{enumerate}

\end{document}